\documentclass[10pt,a4paper,onecolumn]{article}
\usepackage{marginnote}
\usepackage{graphicx}
\usepackage{xcolor}
\usepackage{authblk,etoolbox}
\usepackage{titlesec}
\usepackage{calc}
\usepackage{tikz}
\usepackage{orcidlink}
\usepackage{hyperref}
\hypersetup{colorlinks,breaklinks=true,
            urlcolor=[rgb]{0.0, 0.5, 1.0},
            linkcolor=[rgb]{0.0, 0.5, 1.0}}
\usepackage{caption}
\usepackage{tcolorbox}
\usepackage{amssymb,amsmath}
\usepackage{ifxetex,ifluatex}
\usepackage{seqsplit}
\usepackage{xstring}

\usepackage{float}
\let\origfigure\figure
\let\endorigfigure\endfigure
\renewenvironment{figure}[1][2] {
    \expandafter\origfigure\expandafter[H]
} {
    \endorigfigure
}

\usepackage{fixltx2e} 

\let\textttOrig=\texttt
\def\texttt#1{\expandafter\textttOrig{\seqsplit{#1}}}
\renewcommand{\seqinsert}{\ifmmode
  \allowbreak
  \else\penalty6000\hspace{0pt plus 0.02em}\fi}

\makeatletter
\let\href@Orig=\href
\def\href@Urllike#1#2{\href@Orig{#1}{\begingroup
    \def\Url@String{#2}\Url@FormatString
    \endgroup}}
\def\href@Notdoi#1#2{\def\tempa{#1}\def\tempb{#2}%
  \ifx\tempa\tempb\relax\href@Urllike{#1}{#2}\else
  \href@Orig{#1}{#2}\fi}
\def\href#1#2{%
  \IfBeginWith{#1}{https://doi.org}%
  {\href@Urllike{#1}{#2}}{\href@Notdoi{#1}{#2}}}
\makeatother

\NewDocumentCommand\citeproctext{}{}
\NewDocumentCommand\citeproc{mm}{%
  \hyperlink{cite.#1}{#2}}
\makeatletter
 \let\@cite@ofmt\@firstofone
 \def\@biblabel#1{}
 \def\@cite#1#2{{#1\if@tempswa , #2\fi}}
\makeatother
\newlength{\cslhangindent}
\newlength{\csllabelwidth}
\newenvironment{CSLReferences}[2] 
 {\begin{list}{}{%
  \setlength{\itemindent}{0pt}
  \setlength{\leftmargin}{0pt}
  \setlength{\parsep}{0pt}
  \ifodd #1
   \setlength{\leftmargin}{\cslhangindent}
   \setlength{\itemindent}{-1\cslhangindent}
  \fi
  \setlength{\itemsep}{#2\baselineskip}}}
 {\end{list}}
\usepackage{calc}

\usepackage[top=3.5cm, bottom=3cm, right=1.5cm, left=1.0cm,
            headheight=2.2cm, reversemp, includemp, marginparwidth=4.5cm]{geometry}

\titleformat{\section}
  {\normalfont\sffamily\Large\bfseries}
  {}{0pt}{}
\titleformat{\subsection}
  {\normalfont\sffamily\large\bfseries}
  {}{0pt}{}
\titleformat{\subsubsection}
  {\normalfont\sffamily\bfseries}
  {}{0pt}{}
\titleformat*{\paragraph}
  {\sffamily\normalsize}

\usepackage{fancyhdr}
\makeatletter
\let\ps@plain\ps@fancy
\fancyheadoffset[L]{4.5cm}
\fancyfootoffset[L]{4.5cm}

\definecolor{linky}{rgb}{0.0, 0.5, 1.0}

\newtcolorbox{repobox}
   {colback=red, colframe=red!75!black,
     boxrule=0.5pt, arc=2pt, left=6pt, right=6pt, top=3pt, bottom=3pt}

\newcommand{\ExternalLink}{%
   \tikz[x=1.2ex, y=1.2ex, baseline=-0.05ex]{%
       \begin{scope}[x=1ex, y=1ex]
           \clip (-0.1,-0.1)
               --++ (-0, 1.2)
               --++ (0.6, 0)
               --++ (0, -0.6)
               --++ (0.6, 0)
               --++ (0, -1);
           \path[draw,
               line width = 0.5,
               rounded corners=0.5]
               (0,0) rectangle (1,1);
       \end{scope}
       \path[draw, line width = 0.5] (0.5, 0.5)
           -- (1, 1);
       \path[draw, line width = 0.5] (0.6, 1)
           -- (1, 1) -- (1, 0.6);
       }
   }

\patchcmd{\@maketitle}{center}{flushleft}{}{}
\patchcmd{\@maketitle}{center}{flushleft}{}{}
\patchcmd{\@maketitle}{\LARGE}{\LARGE\sffamily}{}{}
\def\maketitle{{%
  
  \AB@maketitle}}
\makeatletter
\renewcommand\AB@affilsepx{ \protect\Affilfont}
\renewcommand\AB@affilnote[1]{{\bfseries #1}\hspace{3pt}}
\renewcommand{\affil}[2][]%
   {\newaffiltrue\let\AB@blk@and\AB@pand
      \if\relax#1\relax\def\AB@note{\AB@thenote}\else\def\AB@note{#1}%
        \setcounter{Maxaffil}{0}\fi
        \begingroup
        \let\href=\href@Orig
        \let\texttt=\textttOrig
        \let\protect\@unexpandable@protect
        \def\thanks{\protect\thanks}\def\footnote{\protect\footnote}%
        \@temptokena=\expandafter{\AB@authors}%
        {\def\\{\protect\\\protect\Affilfont}\xdef\AB@temp{#2}}%
         \xdef\AB@authors{\the\@temptokena\AB@las\AB@au@str
         \protect\\[\affilsep]\protect\Affilfont\AB@temp}%
         \gdef\AB@las{}\gdef\AB@au@str{}%
        {\def\\{, \ignorespaces}\xdef\AB@temp{#2}}%
        \@temptokena=\expandafter{\AB@affillist}%
        \xdef\AB@affillist{\the\@temptokena \AB@affilsep
          \AB@affilnote{\AB@note}\protect\Affilfont\AB@temp}%
      \endgroup
       \let\AB@affilsep\AB@affilsepx
}
\makeatother

\renewcommand\Affilfont{\sffamily\small\mdseries}
\ifnum 0\ifxetex 1\fi\ifluatex 1\fi=0 
  \usepackage[T1]{fontenc}
  \usepackage[utf8]{inputenc}
  \DeclareUnicodeCharacter{00A0}{~}
  \DeclareUnicodeCharacter{00B4}{'}
  \DeclareUnicodeCharacter{00C1}{\'A}
  \DeclareUnicodeCharacter{00DF}{\ss}
  \DeclareUnicodeCharacter{00E0}{\`a}
  \DeclareUnicodeCharacter{00E1}{\'a}
  \DeclareUnicodeCharacter{00E4}{\"a}
  \DeclareUnicodeCharacter{00E8}{\`e}
  \DeclareUnicodeCharacter{00E9}{\'e}
  \DeclareUnicodeCharacter{00EB}{\"e}
  \DeclareUnicodeCharacter{00ED}{\'i}
  \DeclareUnicodeCharacter{00EE}{\^i}
  \DeclareUnicodeCharacter{00F1}{\~n}
  \DeclareUnicodeCharacter{00F2}{\`o}
  \DeclareUnicodeCharacter{00F3}{\'o}
  \DeclareUnicodeCharacter{00F4}{\^o}
  \DeclareUnicodeCharacter{00F6}{\"o}
  \DeclareUnicodeCharacter{00F9}{\`u}
  \DeclareUnicodeCharacter{00FA}{\'u}
  \DeclareUnicodeCharacter{00FC}{\"u}
  \DeclareUnicodeCharacter{0107}{\'c}
  \DeclareUnicodeCharacter{0130}{\.I}
  \DeclareUnicodeCharacter{0131}{\i}
  \DeclareUnicodeCharacter{0142}{\l}
  \DeclareUnicodeCharacter{0144}{\'n}
  \DeclareUnicodeCharacter{0151}{\H{o}}
  \DeclareUnicodeCharacter{015B}{\'s}
  \DeclareUnicodeCharacter{017C}{\.z}
  \DeclareUnicodeCharacter{2009}{\,}
  \DeclareUnicodeCharacter{2013}{--}
  \DeclareUnicodeCharacter{2019}{'}

\else 
  \ifxetex
    \usepackage{mathspec}
    \usepackage{fontspec}

  \else
    \usepackage{fontspec}
  \fi
  \defaultfontfeatures{Ligatures=TeX,Scale=MatchLowercase}

\fi
\IfFileExists{upquote.sty}{\usepackage{upquote}}{}
\IfFileExists{microtype.sty}{%
\usepackage{microtype}
\UseMicrotypeSet[protrusion]{basicmath} 
}{}

\usepackage{hyperref}
\hypersetup{unicode=true,
            pdftitle={gwsnr: A Python package for efficient signal-to-noise ratio calculations of gravitational waves},
            pdfborder={0 0 0},
            breaklinks=true}
\let\addcontentslineOrig=\addcontentsline
\def\addcontentsline#1#2#3{\bgroup
  \let\texttt=\textttOrig\addcontentslineOrig{#1}{#2}{#3}\egroup}
\let\markbothOrig\markboth
\def\markboth#1#2{\bgroup
  \let\texttt=\textttOrig\markbothOrig{#1}{#2}\egroup}
\let\markrightOrig\markright
\def\markright#1{\bgroup
  \let\texttt=\textttOrig\markrightOrig{#1}\egroup}

\usepackage{graphicx,grffile}
\makeatletter
\def\maxwidth{\ifdim\Gin@nat@width>\linewidth\linewidth\else\Gin@nat@width\fi}
\def\maxheight{\ifdim\Gin@nat@height>\textheight\textheight\else\Gin@nat@height\fi}
\makeatother
\setkeys{Gin}{width=\maxwidth,height=\maxheight,keepaspectratio}
\IfFileExists{parskip.sty}{%
\usepackage{parskip}
}{
\setlength{\parindent}{0pt}
\setlength{\parskip}{6pt plus 2pt minus 1pt}
}
\ifx\paragraph\undefined\else
\let\oldparagraph\paragraph
\renewcommand{\paragraph}[1]{\oldparagraph{#1}\mbox{}}
\fi
\ifx\subparagraph\undefined\else
\let\oldsubparagraph\subparagraph
\renewcommand{\subparagraph}[1]{\oldsubparagraph{#1}\mbox{}}
\fi

\title{\(gwsnr\): A Python package for efficient signal-to-noise ratio
calculations of gravitational waves}

          \author[1]{Hemantakumar
Phurailatpam\,\protect\orcidlink{0000-0002-0471-3724}}
              \author[1]{Otto Akseli
HANNUKSELA\,\protect\orcidlink{0000-0002-3887-7137}}
      
      \affil[1]{Department of Physics, The Chinese University of Hong
Kong, Shatin, New Territories, Hong Kong}
  \date{\vspace{-7ex}}

\begin{document}
\maketitle

\marginpar{

  \begin{flushleft}
  \sffamily\small

  {\bfseries DOI:} \href{https://doi.org/10.xxxxxx/draft}{\color{linky}{10.xxxxxx/draft}}

  \vspace{2mm}

  {\bfseries Software}
  \begin{itemize}
    \setlength\itemsep{0em}
    \item \href{https://github.com/openjournals/joss-reviews/issues}{\color{linky}{Review}} \ExternalLink
    \item \href{https://github.com/hemantaph/gwsnr}{\color{linky}{Repository}} \ExternalLink
    \item \href{https://doi.org/10.5281/zenodo.17803640}{\color{linky}{Archive}} \ExternalLink
  \end{itemize}

  \vspace{2mm}

  \par\noindent\hrulefill\par

  \vspace{2mm}

  {\bfseries Editor:} \href{https://github.com/openjournals}{Open Journals} \ExternalLink \\
  \vspace{1mm}
    {\bfseries Reviewers:}
  \begin{itemize}
  \setlength\itemsep{0em}
    \item \href{https://github.com/openjournals}{@openjournals}
    \end{itemize}
    \vspace{2mm}

  {\bfseries Submitted:} 12 January 2024\\
  {\bfseries Published:} unpublished

  \vspace{2mm}
  {\bfseries License}\\
  Authors of papers retain copyright and release the work under a Creative Commons Attribution 4.0 International License (\href{http://creativecommons.org/licenses/by/4.0/}{\color{linky}{CC BY 4.0}}).

  \end{flushleft}
}

\vspace{1.5em}

\section{Summary}\label{summary}

Gravitational waves (GWs), ripples in spacetime predicted by Einstein's
theory of General Relativity, have revolutionized astrophysics since
their first detection in 2015. Emitted by cataclysmic events such as
mergers of binary black holes (BBHs), binary neutron stars (BNSs), and
black hole-neutron star pairs (BH-NSs), these waves provide a unique
window into the cosmos. A central quantity in GW analysis is the
Signal-to-Noise Ratio (SNR), which measures the strength of a GW signal
relative to the background noise in detectors such as LIGO
(\citeproc{ref-Abbott:2020}{Abbott et al., 2020};
\citeproc{ref-Buikema:2020}{Buikema et al., 2020};
\citeproc{ref-LIGO:2015}{The LIGO Scientific Collaboration et al.,
2015}), Virgo (\citeproc{ref-VIRGO:2015}{F. Acernese et al., 2014};
\citeproc{ref-VIRGO:2019}{F. Acernese et al., 2019}), and KAGRA
(\citeproc{ref-Akutsu:2020}{Akutsu et al., 2020};
\citeproc{ref-Aso:2013}{Aso et al., 2013}). While real detections are
established using a False-Alarm Rate (FAR) threshold, under stationary
Gaussian noise assumptions the condition that the SNR exceeds a chosen
threshold can serve as a practical proxy
(\citeproc{ref-Essick:2023}{Essick, 2023};
\citeproc{ref-Essick:2024}{Essick \& Fishbach, 2024}). This proxy is
especially useful in simulations of detectable events and in studies
aimed at extracting astrophysical information
(\citeproc{ref-Abbott:2016:detection}{Abbott, B. P. et al., 2016}).

We introduce \(gwsnr\), a Python package explicitly designed to
efficiently compute the SNR and estimate the probability of detection
for large simulated populations of GW events. By accelerating these core
calculations, \(gwsnr\) makes massive population-level astrophysical
simulations computationally viable.

\section{Statement of need}\label{statement-of-need}

Applications such as population simulations for rate estimation
(\citeproc{ref-Abbott:2016:rates}{Abbott et al., 2016}) and hierarchical
Bayesian inference with selection effects
(\citeproc{ref-Essick:2024}{Essick \& Fishbach, 2024};
\citeproc{ref-Thrane:2019}{Thrane \& Talbot, 2019}) require repeated and
efficient computation of the probability of detection (\(P_{\rm det}\)).
This probability is generally derived from the SNR. However, traditional
approaches that rely on noise-weighted inner products for SNR evaluation
are computationally demanding and often impractical for large-scale
analyses (\citeproc{ref-Gerosa:2020}{Gerosa et al., 2020};
\citeproc{ref-Taylor:2018}{Taylor \& Gerosa, 2018}).

The \(gwsnr\) package resolves this computational bottleneck by rapidly
evaluating the
\href{https://gwsnr.hemantaph.com/detectionstatistics.html\#defining-optimal-snr}{optimal-SNR}
(\(\rho_{\rm opt}\)) and \(P_{\rm det}\). The target audience includes
LIGO-Virgo-KAGRA (LVK) researchers and astrophysicists working on
population studies, selection-effect modeling, and astrophysical rate
estimation. For massive simulations, \(gwsnr\) estimates \(P_{\rm det}\)
by evaluating \(\rho_{\rm opt}\) against a detection-statistic threshold
under flexible detector, waveform, and population settings. It also
enables statistical modeling of the noise-realized
\href{https://gwsnr.hemantaph.com/detectionstatistics.html\#defining-match-filter-snr}{observed/matched-filter
SNR} (\(\rho_{\rm obs}\)) based on \(\rho_{\rm opt}\) under the
stationary Gaussian noise assumption. By accelerating \(\rho_{\rm opt}\)
evaluation with specialized numerical methods, \(gwsnr\) makes rapid
calculations possible under diverse astrophysical configurations that
would otherwise be prohibitive. Although intended for a variety of
research purposes, \(gwsnr\) operates effectively as a specialized
detectability calculator and is heavily utilized by the \texttt{ler}
(\citeproc{ref-ler:2024}{Phurailatpam et al., 2024}) software for
simulating detectable strongly lensed GW events and calculating their
rates (\citeproc{ref-More:2025}{More \& Phurailatpam, 2025};
\citeproc{ref-Leo:2024}{Ng et al., 2024}).

\section{State of the field}\label{state-of-the-field}

In GW data analysis, commonly used packages such as \texttt{Bilby}
(\citeproc{ref-Ashton:2019}{Ashton et al., 2019}), \texttt{PyCBC}
(\citeproc{ref-Usman:2016}{Usman et al., 2016}), and \texttt{GstLAL}
(\citeproc{ref-gstlal:2020}{Cannon et al., 2020}), operating alongside
\texttt{LALSuite} (\citeproc{ref-lalsuite:2018}{LIGO Scientific
Collaboration et al., 2018}), provide high-precision tools for Bayesian
inference, matched-filter searches, signal processing, and waveform
generation. While widely adopted, these packages are not primarily
designed for the fast, repeated evaluation of \(\rho_{\rm opt}\) and
\(P_{\rm det}\) across millions of simulated events, making direct
noise-weighted inner-product calculations a significant bottleneck. To
address this, related studies have developed machine-learning methods to
estimate detectability. Gerosa et al. (\citeproc{ref-Gerosa:2020}{2020})
trained classifiers using optimal-SNR thresholds, while Callister et al.
(\citeproc{ref-Callister:2024}{2024}) trained an emulator directly from
search-pipeline detections defined by FAR criteria. Similarly,
Chapman-Bird et al. (\citeproc{ref-ChapmanBird:2023}{2023}) implemented
an SNR-based selection model for extreme mass ratio inspirals using
neural networks. While these machine-learning methods provide rapid
evaluation, their applicability strictly depends on specific training
data, detector configurations, and waveform models. Furthermore, errors
concentrate near the sharp detection boundary, risking the
misclassification of marginal events (\citeproc{ref-Gerosa:2020}{Gerosa
et al., 2020}). Modifying detector noise curves or waveform families in
these emulators requires computationally expensive dataset regeneration
and complete model retraining.

The \(gwsnr\) package addresses these limitations by providing a modular
framework dedicated strictly to efficient detectability calculations.
The build-versus-contribute justification is founded on the necessity
for an adaptable library focused solely on repeated SNR and
\(P_{\rm det}\) evaluations with user-controlled parameters. Unlike
broad inference packages or rigid machine-learning emulators, \(gwsnr\)
explicitly computes \(\rho_{\rm opt}\) through accelerated numerical
pathways while allowing \(\rho_{\rm obs}\) to be modeled statistically.
The detection threshold can also be estimated from injection catalogues
using empirical sensitivity information
(\citeproc{ref-Essick:2023}{Essick, 2023}). This framework allows
researchers to seamlessly switch between partial-scaling interpolation,
artificial neural network approximations, and full multiprocessing
inner-product integration depending on their specific accuracy and speed
requirements. By interfacing with established C-based waveform
infrastructure such as \texttt{LALSuite} and JAX-based tools such as
\texttt{ripplegw} (\citeproc{ref-Edwards:2023}{Edwards et al., 2024}),
\(gwsnr\) fills a unique scholarly role as an adaptable high-speed
detectability engine for broader simulation pipelines.

\section{Software design}\label{software-design}

The architecture of \(gwsnr\) is driven by the trade-off between
accuracy, speed, and waveform generality. Because a single evaluation
method is insufficient for diverse astrophysical studies, \(gwsnr\)
implements multiple calculation paths under a unified interface
parameterized by the detector power spectral density (PSD), antenna
response, waveform model, detector network, source parameters, and
detection threshold.

The baseline calculation executes the standard frequency-domain
\href{https://gwsnr.hemantaph.com/innerproduct.html\#noise-weighted-inner-product-method}{inner
product} (\citeproc{ref-Allen:2012}{Allen et al., 2012}) given by the
equation

\[
\langle a | b \rangle =
4 \Re \int_{f_{\min}}^{f_{\max}}
\frac{\tilde{a}(f)\tilde{b}^*(f)}{S_n(f)} df
\]

where \(S_n(f)\) is the detector PSD. The optimal SNR is
\(\rho=\sqrt{\langle h|h\rangle}\). For the two polarizations \(h_+\)
and \(h_\times\), and given antenna patterns of the detector (\(F_+\),
\(F_\times\)), the optimal SNR is calculated as

\[
\rho =
\sqrt{
F_+^2 \langle \tilde{h}_+|\tilde{h}_+\rangle
+
F_\times^2 \langle \tilde{h}_\times|\tilde{h}_\times\rangle  
},  
\]

where I have assumed orthogonality between \(h_+\) and \(h_\times\).

This path is prioritized for its generality and supports spin-precessing
systems alongside waveforms with subdominant modes. \(gwsnr\)
accelerates this conventional approach using Python multiprocessing,
Just-In-Time (JIT) compilation via \texttt{numba.njit}
(\citeproc{ref-numpy:2022}{NumPy Community, 2022}) for antenna-pattern
generation, and optional \texttt{JAX} (\citeproc{ref-jax:2018}{Bradbury
et al., 2018}) backends integrated with \texttt{ripplegw}
(\citeproc{ref-Edwards:2023}{Edwards et al., 2024}). With \(n\)
allocated processes,
\href{https://gwsnr.hemantaph.com/performancesummary.html\#inner-product-performance}{multiprocessing
delivers speedups} approaching a factor of \(n\), with minimal overhead.

For non-spinning and aligned-spin binaries, \(gwsnr\) implements a
fundamentally faster
\href{https://gwsnr.hemantaph.com/interpolation.html\#the-partial-scaling-interpolation-method}{partial-scaling
interpolation} method based on the FINDCHIRP scaling
(\citeproc{ref-Allen:2012}{Allen et al., 2012}). This architectural
choice avoids repeated inner-product integrations by isolating the
computationally expensive mass-dependent and spin-dependent components.
The method precomputes the partial-scaled SNR defined as

\[
\rho_{1/2}
=
\frac{D_\mathrm{eff}}{\mathcal{M}^{5/6}}
\rho_{\rm opt}
\]

where \(\mathcal{M}\) is the chirp mass and \(D_{\rm eff}\) is the
effective distance. The quantity \(\rho_{1/2}\) is stored on
multidimensional irregular grids using local cubic-Hermite splines. New
SNR values are rapidly recovered by interpolation and rescaling using
the equation

\[
\rho_{\rm opt}
=
\rho_{1/2}
\frac{\mathcal{M}^{5/6}}{D_\mathrm{eff}} .
\]

The full \(\rho_{\rm opt}\) generation is wrapped in JIT-compiled
functions for parallel execution on CPUs (via the \texttt{numba}
backend) and on supported GPUs, including Apple Silicon (via the
\texttt{MLX} backend (\citeproc{ref-mlx:2023}{Hannun et al., 2023})) and
Nvidia GPUs (via the \texttt{JAX} backend). For randomly sampled
parameters, this method achieves accuracy exceeding \(99.5\) compared
with traditional noise-weighted inner-product calculations using
\texttt{Bilby} (\citeproc{ref-Ashton:2019}{Ashton et al., 2019}), and
processes up to one million \(\rho_{\rm opt}\) values in \(\sim 200\)
milliseconds (max case) on GPU backends, with
\href{https://gwsnr.hemantaph.com/performancesummary.html\#interpolation-performance}{speedups}
of \(\sim 5{,}000\times\) relative to \texttt{Bilby}. The interpolation
grid must be generated once beforehand (\(\lesssim 1\) min with default
settings) and is stored in JSON files for reuse; \texttt{numba.njit}
compilation overhead is negligible thereafter.

As a supplementary path, \(gwsnr\) includes
\href{https://gwsnr.hemantaph.com/ann.html\#ann-based-pdet-estimation}{artificial
neural network (ANN) estimation} using \texttt{tensorflow}
(\citeproc{ref-tensorflow:2015}{Abadi et al., 2015}) and
\texttt{scikit-learn} (\citeproc{ref-scikitlearn:2011}{Pedregosa et al.,
2011}). This design is applied to complex waveform settings where direct
partial-scaling interpolation is geometrically impractical. The model
strategically uses partial-scaled SNR quantities to reduce the input
dimensionality from fifteen parameters down to five. Users are also
provided with the tools to train their own models for different
detector, waveform, and population settings.

To manage the boundary classification errors inherent to approximate
methods, \(gwsnr\) introduces a
\href{https://gwsnr.hemantaph.com/hybrid.html\#hybrid-strategy-for-spin-precessing-systems}{hybrid
SNR recalculation} workflow. It estimates initial SNRs using partial
scaling or ANN prediction, isolates events near the detection threshold
\(\rho_{\rm th}\), and exactly recalculates those specific systems using
the direct inner-product method. This design preserves the speed of
interpolation while guaranteeing reliability near the critical detection
boundary where small SNR errors can directly affect \(P_{\rm det}\).

The
\href{https://gwsnr.hemantaph.com/probabilityofdetection.html\#probability-of-detection-calculation}{probability
of detection} is evaluated by applying a threshold to either
\(\rho_{\rm opt}\) or \(\rho_{\rm obs}\), denoted by
\(\rho_{\rm opt,th}\) and \(\rho_{\rm obs,th}\), respectively. For a
deterministic threshold on a chosen SNR quantity \(\rho\),

\[
P_{\rm det}
=
P({\rm det}\mid\vec{\theta})
=
\Theta(\rho-\rho_{\rm th}),
\]

where \(\vec{\theta}\) denotes the GW parameters and \(\Theta\) is the
Heaviside step function. If \(\rho_{\rm obs}\) is treated as a
noise-dependent random variable, the probability is averaged over noise
realizations,

\[
P({\rm det}\mid\vec{\theta})
=
P(\rho_{\rm obs}>\rho_{\rm obs,th}\mid\vec{\theta})
=
1-F_{\rho_{\rm obs}}(\rho_{\rm obs,th}\mid\vec{\theta}),
\]

where \(F_{\rho_{\rm obs}}\) is the cumulative distribution function
under the assumed noise model. Under the stationary Gaussian noise
approximation, \(gwsnr\) supports two models for \(\rho_{\rm obs}\). The
first treats \(\rho_{\rm obs}\) as a normal random variable centred on
\(\rho_{\rm opt}\) with unit variance (\citeproc{ref-Abbott:2019}{Abbott
et al., 2019}; \citeproc{ref-Fishbach:2020}{Fishbach et al., 2020}). The
second treats \(\rho_{\rm obs}^2\) as a non-central \(\chi^2\) variable.
Using the convention of \texttt{scipy.stats.ncx2}
(\citeproc{ref-scipy:2020}{Virtanen et al., 2020}),

\[
\rho_{\rm obs}^2
\sim
\chi^2_{\rm nc}
\left(
k=2,
\lambda=\rho_{\rm opt}^2
\right)
\]

for a single detector, and

\[
\rho_{\rm obs,net}^2
\sim
\chi^2_{\rm nc}
\left(
k=2N,
\lambda=\rho_{\rm opt,net}^2
\right)
\]

for a network of \(N\) detectors.

\(gwsnr\) also supports user-defined thresholds and
\href{https://gwsnr.hemantaph.com/examples/threshold.html\#SNR-Threshold-Finder-Example}{provides
tools} for estimating thresholds from injection catalogues, following
Essick (\citeproc{ref-Essick:2023}{2023}). The current implementation
treats this threshold as parameter-independent. Parameter-dependent
thresholds and the corresponding \(P_{\rm det}\) calculation are left
for future development.

Finally, \(gwsnr\) calculates the
\href{https://gwsnr.hemantaph.com/horizondistance.html\#horizon-distance}{horizon
distance} (\(D_{\rm hor}\)), representing the maximum distance at which
an optimally oriented source can be detected
(\citeproc{ref-Allen:2012}{Allen et al., 2012}). The analytical path
simply rescales a known effective distance using

\[
D_{\rm hor}
=
\frac{\rho_{\rm opt}}{\rho_{\rm opt,th}} D_{\rm eff}.
\]

The alternative numerical method maximizes the SNR over sky location and
solves for the luminosity distance \(d_L\) where

\[
\rho(d_L) - \rho_{\rm opt,th} = 0 .
\]

\section{Research impact statement}\label{research-impact-statement}

The \(gwsnr\) package provides the computational speed and accuracy
required for detectability calculations within large simulated
populations of compact binary mergers. By making massive SNR evaluations
computationally viable, the software actively supports astrophysical
\href{https://ler.hemantaph.com/examples/LeR_custom_functions.html}{rate
estimation},
\href{https://gwsnr.hemantaph.com/examples/horizon_distance.html}{detector-sensitivity}
studies, and
\href{https://ler.hemantaph.com/examples/selection_function.html}{selection-effect
modeling} in hierarchical Bayesian inference. A primary realized impact
of \(gwsnr\) is its functional integration as the core detectability
calculator within the \texttt{ler} software package
(\citeproc{ref-ler:2024}{Phurailatpam et al., 2024}). In this capacity,
it identifies detectable unlensed and strongly lensed gravitational-wave
events to calculate their expected occurrence rates, directly
facilitating population-level lensing analyses in published literature
(\citeproc{ref-Janquart:2023}{Janquart et al., 2023};
\citeproc{ref-More:2025}{More \& Phurailatpam, 2025};
\citeproc{ref-Leo:2024}{Ng et al., 2024}).

The community readiness of \(gwsnr\) is demonstrated by its formal peer
review within the LIGO-Virgo-KAGRA Scientific Collaboration
(\citeproc{ref-gwsnrlvkpnpreview:2024}{LIGO-Virgo-KAGRA Collaboration,
2024}). The software maintains active support through
\href{https://github.com/hemantaph/gwsnr/issues}{GitHub issues} and
\href{https://chat.ligo.org/}{collaboration communication channels}. The
package is publicly accessible on the Python Package Index where it has
recorded over 400 downloads (\citeproc{ref-gwsnrpypi:2026}{Phurailatpam
\& Hannuksela, 2026b}). Comprehensive documentation provides the
underlying theory, tutorials, and benchmarks necessary to reproduce all
core calculations (\citeproc{ref-gwsnrdocs:2026}{Phurailatpam \&
Hannuksela, 2026a}). These factors provide specific and compelling
evidence that \(gwsnr\) successfully supports independent simulation
pipelines tailored with custom detector, waveform, and population
parameters.

\section{AI usage disclosure}\label{ai-usage-disclosure}

No generative AI tools were used to develop the software, write this
manuscript, or prepare the accompanying materials.

\section{Acknowledgements}\label{acknowledgements}

Hemantakumar Phurailatpam acknowledges the Department of Physics at The
Chinese University of Hong Kong for the Postgraduate Studentship that
facilitated this research. Hemantakumar Phurailatpam and Otto A.
Hannuksela acknowledge support from the Research Grants Council of Hong
Kong, Project Nos. CUHK 14304622 and 14307923, the start-up grant from
The Chinese University of Hong Kong, and the Direct Grant for Research
from the Research Committee of The Chinese University of Hong Kong. The
authors also thank the LIGO Laboratory for computational resources,
supported by National Science Foundation Grants No.~PHY-0757058 and
No.~PHY-0823459.

\section*{References}\label{references}
\addcontentsline{toc}{section}{References}

\phantomsection\label{refs}

\end{document}